\documentclass[conference]{IEEEtran}
\IEEEoverridecommandlockouts

\usepackage{cite}
\usepackage{amsmath,amssymb,amsfonts}
\usepackage{algorithmic}
\usepackage{graphicx}
\usepackage{textcomp}
\usepackage{xcolor}
\usepackage{orcidlink}
\usepackage{tcolorbox}
\usepackage{todonotes}
\usepackage{booktabs}
\usepackage{hyperref}
\usepackage{xurl}
\usepackage{array}
\usepackage{tcolorbox}
\usepackage{fontawesome5}
\usepackage{makecell}
\usepackage{array, multirow, boldline}
\usepackage{caption}

\def\BibTeX{{\rm B\kern-.05em{\sc i\kern-.025em b}\kern-.08em
    T\kern-.1667em\lower.7ex\hbox{E}\kern-.125emX}}

\usepackage[T1]{fontenc}
\begin{document}

\title{LLMs for Generation of Architectural Components: An Exploratory Empirical Study in the Serverless World}

\author{\IEEEauthorblockN{Shrikara Arun* 
\orcidlink{0009-0005-8543-6130}
} 
\IEEEauthorblockA{
\textit{Software Engineering Research Centre} \\
\textit{IIIT Hyderabad, India}\\
\textit{shrikara.a@students.iiit.ac.in}}
\thanks{*These authors contributed equally.}
\and
\IEEEauthorblockN{Meghana Tedla* 
\orcidlink{0009-0000-8540-8790}
}
\IEEEauthorblockA{
\textit{Software Engineering Research Centre} \\
\textit{IIIT Hyderabad, India}\\
\textit{meghana.tedla@students.iiit.ac.in}}
\and
\IEEEauthorblockN{Karthik Vaidhyanathan 
\orcidlink{0000-0003-2317-6175}
}
\IEEEauthorblockA{
\textit{Software Engineering Research Centre} \\
\textit{IIIT Hyderabad, India}\\
\textit{karthik.vaidhyanathan@iiit.ac.in}}
 }
\maketitle

\newcommand{\magenta}[1]{\textcolor{magenta}{#1}}
\newcommand{\blue}[1]{\textcolor{blue}{#1}}
\newcommand{\Human}{\stackon{\faUser}{\textcolor{green}{\faCheck}}}
\newcommand{\NoHuman}{\stackon{\faUser}{\textcolor{red}{\faTimes}}}

\begin{abstract}
Recently, the exponential growth in capability and pervasiveness of Large Language Models (LLMs) has led to significant work done in the field of code generation. However, this generation has been limited to code snippets. Going one step further, our desideratum is to automatically generate architectural components. This would not only speed up development time, but would also enable us to eventually completely skip the development phase, moving directly from design decisions to deployment. To this end, we conduct an exploratory study on the capability of LLMs to generate architectural components for Functions as a Service (FaaS), commonly known as serverless functions. The small size of their architectural components make this architectural style amenable for generation using current LLMs compared to other styles like monoliths and microservices. We perform the study by systematically selecting open source serverless repositories, masking a serverless function and utilizing state of the art LLMs provided with varying levels of context information about the overall system to generate the masked function. We evaluate correctness through existing tests present in the repositories and use metrics from the Software Engineering (SE) and Natural Language Processing (NLP) domains to evaluate code quality and the degree of similarity between human and LLM generated code respectively. Along with our findings, we also present a discussion on the path forward for using GenAI in architectural component generation.
\end{abstract}


\vspace{-12pt}
\section{Introduction}
Ever since the inception of the field of Software Architecture (SA)~\cite{garlan1993_component_connector}, one of the goals has been to automate/semi-automate the generation of executable systems from architecture descriptions as this would increase compliance, promote traceability, etc. Over the years the SA community has been working towards realizing this goal through defining Architecture Description Languages (ADLs) and Domain Specific Languages (DSLs)~\cite{mehmood2013aspect,malavolta2012industry,aadl-robotics} and further developing approaches for code generations using model transformations. However, their application to practice has been limited due to the steep learning curve, lack of extensibility, support for tooling, etc~\cite{malavolta2012industry}.  On the other hand, with recent advancements in AI, Large Language Models (LLMs) are moving us ever closer to a world of increased automation, with applications across multiple Software Engineering (SE) tasks, as described by Hou et al. \cite{davidlo_llm_slr}. They have been used for software development, maintenance, requirements engineering, and more, with code generation and program repair being the most common applications \cite{davidlo_llm_slr}. There have also been several commercial tools such as ChatGPT, GitHub Copilot, and Cursor. However, this code generation has been in the context of generating low level code snippets, with the generation of software architecture components using LLMs being an unexplored space. 

To this end, we conduct an exploratory empirical study on the capability of LLMs to generate architectural components in the context of Functions-as-a-Service (FaaS), commonly referred to as serverless functions. FaaS supports event-driven architectural style and enables easy development due to the abstractions provided by the cloud provider, who manages the infrastructure for running the basic units of FaaS, called serverless functions. We choose serverless functions primarily due to the small size of their architectural component, as opposed to microservices or monoliths, where a single component may consist of thousands or even millions of lines of code. We believe that this can provide a first step when evaluating the architectural component generation capabilities of LLMs. We emphasize that the architectural components we deal with in our study are serverless functions, and we refer to the architectural component of FaaS as serverless functions in the remainder of this paper. As part of our study, we utilize 3 kinds of prompts containing information at different levels of abstraction, systematically select 4 open-source serverless repositories and 5 code-generation LLMs, generating a total of 145 serverless functions that we evaluate for functionality and code quality both with and without human intervention. 
The code and data for our study is available publicly. \footnote{Code and data available at: \url{https://doi.org/10.5281/zenodo.14539782}}


The remainder of this paper is structured as follows: Section \ref{background} provides background information about Serverless Functions, LLMs and some prompting methods. Section \ref{related-work} describes related work. Section \ref{study-design} describes our research questions and the design of our study. Section \ref{results} presents results, which are discussed in Section \ref{discussion} along with a look into a possible future for GenAI for Software Architecture and the threats to validity of our study. Finally, Section \ref{conclusion} presents our conclusion and future work.

\section{Background} \label{background}
This section provides an overview of the concepts dealt with in this study, including serverless functions, LLMs for general use and for code generation and methods to improve performance of an LLM on a task.

\vspace*{-2.5mm}

\subsection{Function-as-a-Service (FaaS) (Serverless Functions)} \label{backg:faas}

Function-as-a-Service (FaaS) is one of the most popular forms of the cloud computing paradigm called serverless computing\cite{wen2023rise_serverlessslr}. It is widely adopted across domains due to its nature of being \textit{ephemeral}, \textit{event driven}, and \textit{elastic}. Unlike traditional cloud computing paradigms, in FaaS, developers only write business logic in the form of fine-grained \textit{functions}, which are deployed onto a cloud platform, where the cloud provider handles the hassle of managing and maintaining infrastructure, execution environment, which is abstracted away from the developer. FaaS functions are \textit{event driven} as they only run when triggered by an event, such as a HTTP request, an update in a database, or a message arriving on a message queue. They are \textit{ephemeral} since they only run for a short amount of time, after which they are descheduled by the cloud provider. They are also \textit{elastic}, since they can scale automatically based on load. This also leads to potential cost savings by charging only for the time functions run. These properties and the multitude of languages supported makes FaaS applications quite easy to develop. Through slight abuse of notation, we refer to serverless functions and functions interchangeably in the remainder of this paper.

\subsection{Large Language Models and Code Generation}\label{backg:llm}
Large Language Models (LLMs) are probabilistic machine learning models built on the Transformer architecture \cite{vaswani2017attention} that can mimic human language use. They are composed of billions of parameters and are trained on substantial content sourced from the internet. Though they can be trained on and used for other tasks, paradigmatic LLMs are \textit{generative auto-regressive models}, meaning that they sequentially generate the next token in a sequence of tokens. Tokens are the units into which text is broken down when processed and generated by LLMs. They may correspond to an entire word or to a part of a word \cite{sennrich-etal-2016-bpe}. The maximum number of tokens that an LLM can process is called its \textit{context length} or \textit{context window}, and this limits the amount of information that can be provided in the input \textit{prompt} by the user.  

The coding ability of LLMs was improved by the creation of datasets like HumanEval \cite{chen2021evaluating} and Mostly Basic Python Problems (MBPP) \cite{austin2021mbpp} on which LLMs have been \textit{fine-tuned}. Fine-tuning involves adapting a pre-trained model to perform better on a specific task.

\subsection{Zero-Shot Prompting and Few-Shot Prompting}\label{backg:llm_prompting}
Unlike fine-tuning, zero-shot and few-shot prompting do not require 
training of the language model or
a dataset to fine-tune on. Zero-shot prompting involves directly making a model perform a task without examples, whereas few-shot prompting provides some examples in context. Few-shot prompting is also referred to as in-context learning, as it improves the performance of the model on a task without requiring updates to the model's parameters \cite{jurafsky_martin}. This significantly improves model performance on the task, especially for larger models, as shown by \cite{brown2020language}. In our work, an instance of zero-shot prompting is to simply provide context (the content of which is discussed in \ref{study:generation}) and request a serverless function to be generated. For few-shot prompts, we additionally provide some serverless functions from the repository along with their description to the LLM and ask for the function corresponding to a description whose code is not specified.
\vspace*{-1mm}

\vspace{-7pt}
\section{Related Work} \label{related-work}
Over the years, a lot of promising work has been done in specifying software architecture using Architecture Description Languages (ADL) and further supporting the analysis and code generation of components using model transformations \cite{brun2008code, malavolta2012industry}. In particular, an approach for the generation of source code in the Go programming language from $\pi$-ADL was proposed by Cavalcante et al. \cite{cavalcante2014architectureadl}. Further, an ADL for architecture modeling, code generation, and simulation of Cyber-Physical-Systems was proposed by Sharaf et al. \cite{muccini2017caps,muccini_arduino}. Bardaro et al. in their recent work \cite{aadl-robotics} make use of AADL for modeling and further code generation in the robotics domain. 
However, to the best of our knowledge, not much work has been done in the serverless domain. Tankov et al.\cite{kotless} proposes Kotless which aims to simplify serverless development through the use of a DSL and a plugin to generate deployment code. However, the developer must still develop the application logic and code. There have been some works in using DSLs for supporting the specification and generation of microservices. Rademacher et al.\cite{lemma-microservices}
propose an extensible approach to generate adaptable microservice code and deployment specifications from models developed using the modeling language, LEMMA. Further, Suljkanovi{\'c} et al.\cite{silveria} proposed Silveria, a DSL that allows users to model the architecture of microservice-based systems and further generate executable code using model transformations. However as pointed out by Malavolta et al.\cite{malavolta2012industry}, ADLs in general have a higher learning curve which often hinders practical adoption. 

Significant work has been done on using LLMs for code generation, including \cite{bareiss2022codegen, gong2023intendedcodegen, chen2023improvingcodegen, wang2023review_llmcodegeneration, jiang2024surveylargelanguagemodels}. There also exist several tools such as GitHub Copilot and the Cursor IDE. However, these focus on the generation of code snippets, such as classes or methods, and not at an architectural component level. 

While Eskandani et al. \cite{eskandani2024towards_aisystems} describe the application design of serverless systems using GenAI as an open research direction, it is through the lens of selecting an optimal pattern based on requirements. We seek to evaluate the capability of LLMs in generating architectural components by choosing serverless as the architectural style. To this end, we also conduct evaluations on the functionality, code quality, and similarity between human-written and generated code.

\section{Study Design} \label{study-design}
\vspace{-2mm}
\subsection{Goal}
This study aims to evaluate the degree to which LLMs are able to generate software architecture components. The degree here refers to both the functional correctness and quality of code. Using the Goal-Question-Metric approach described in \cite{caldiera1994goal}, we formalize our goal to:
\\
\textbf{analyze} the effectiveness of LLMs
\textbf{for the purpose of} generating software architecture components
\textbf{with respect to} automatic software architectural component generation
\textbf{from the viewpoint of} software architects and developers
\textbf{in the context of} the Function-as-a-Service (FaaS) architectural style

\subsection{Research Questions} \label{rqs}
\textbf{RQ$_{1}$}: \textit{Can LLMs generate functional serverless functions?} 
\begin{itemize}
    \item \textbf{RQ$_{1.1}$}: \textit{Can LLMs generate functional serverless functions without human intervention?} 
    \item \textbf{RQ$_{1.2}$}: \textit{Can LLMs generate functional serverless functions with minimal human intervention?}
\end{itemize} 
In this RQ, we aim to evaluate whether the architectural components generated by LLMs are usable, to the extent that they satisfy tests defined in their containing repositories. To this end, we measure the outcome both when we directly use the generated serverless function, and after fixing minor errors described in Table \ref{table:llm_error_table}.

\textbf{RQ$_{2}$}: \textit{How does the code quality of LLM-written serverless functions compare to human-written code?} \\
In this RQ, we move beyond functionality to evaluate the code quality of the architectural component generated by the LLM. We are also interested in measuring how similar generated serverless functions are to their human written counterparts.


\subsection{Experiment Workflow}
\begin{figure*}[htb!]
    \centering
    \includegraphics[width=0.85\textwidth]{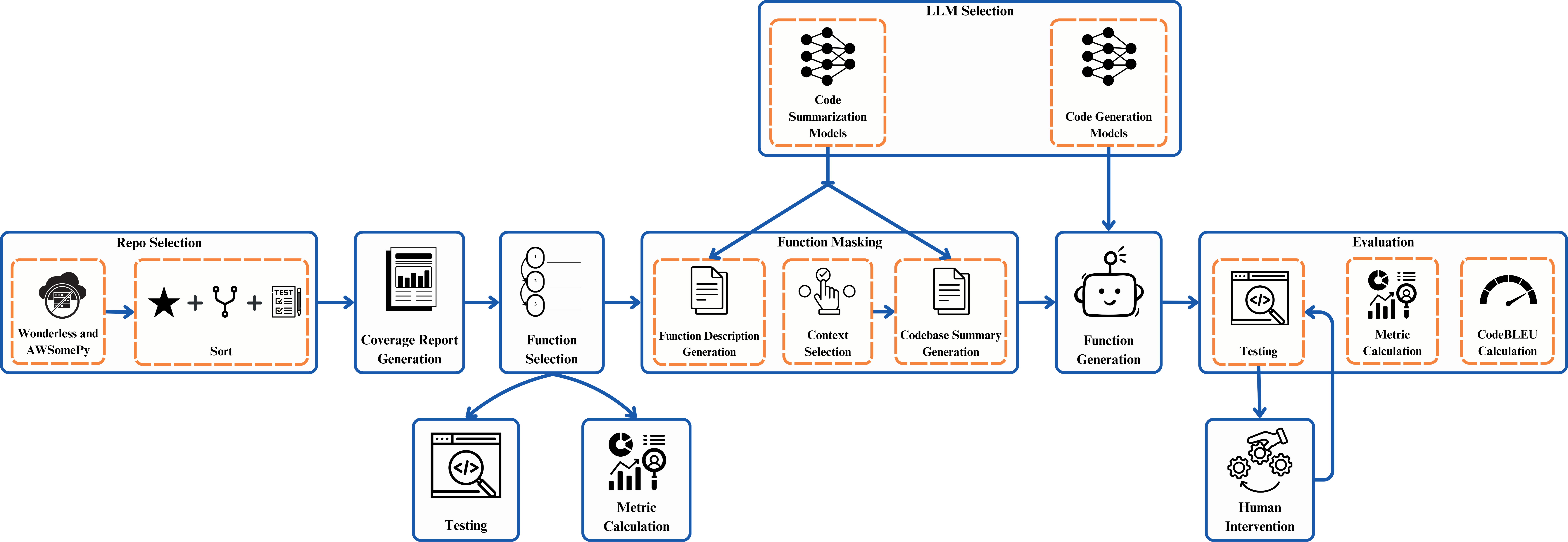}
    \caption{Study Design}
    \label{fig:study-design}
\vspace{-13pt}
\end{figure*}

In this subsection, we first explain Function Masking, an adaptation of Masked Language Modeling from \cite{kenton2019bert}, which we use to generate architectural components. Then, we describe each stage of the workflow shown in Figure \ref{fig:study-design}.
\subsubsection{Function Masking}
BERT \cite{kenton2019bert} was trained on the Masked Language Modeling (MLM) task, where tokens were randomly masked from the input, with the objective to predict the masked token based on its context (both to the left and right). This task is useful and allows for self-supervised learning \cite{liu_selfsupervised}. For example, in the sentence "The tall boy goes to the market", if "boy" is chosen to be masked, the model sees the input "The tall [MASK] goes to the market", where "[MASK]" is the special mask token, and must output "boy". We extend this concept to the serverless world through Function Masking. Given a serverless repository with multiple serverless functions, we mask one function and use the LLM to recreate the masked function, using different types and amounts of context, as explained in \ref{study:context}. 

\subsubsection{LLM Selection} \label{study:llm-selection}
An extensive number of LLMs are available for use, and we select some for our study using leaderboards published in literature. We require two LLMs, one which will be used for summarizing the existing codebase, as described in \ref{study:context}, and one for generating the new serverless function, as described in \ref{study:generation}. The reason for choosing two distinct models is two-fold: \\
    Primarily, the tasks these two models will be performing are different. The summarization model should be able to create an abstractive summary from the code given to it as context along with capturing the architectural information. This necessitates that it has a long context length, so that the input prompt can accommodate all the code needed to create the summary. Since a dedicated benchmark does not exist for this task, we use the ChatBot Arena\footnote{Leaderboard available at \url{https://lmarena.ai/}} \cite{chiang2024chatbotarenaopenplatform}, which evaluates human preference. \\ 
    However, the code generation model performs a coding task for which there are existing benchmarks (such as HumanEval \cite{chen2021evaluating} and MBPP \cite{austin2021mbpp}) and fine-tuned models (such as DeepSeekCoder, CodeQwen). We select models from the top 10 of the EvalPlus leaderboard\footnote{Leaderboard available at \url{https://evalplus.github.io/leaderboard.html}} \cite{liu2024code_evalplus}, while ensuring diversity across size and availability. Both EvalPlus and Chat-Bot Arena are evolving leaderboards, and the models selected were in the top 10 at the time of conducting the study.   \\
    Secondly, using the same model may propagate biases, such as which part of the codebase is important. This avoids the possibility of the model creating a summary that can only be fully deciphered by itself, and not by other models and humans.

We selected Gemini-1.5-Pro\cite{team2024gemini} for codebase summarization.
The models selected for code generation are described in Table \ref{tab:model-info} below. Refer \ref{backg:llm} for an explanation on "Number of Parameters" and "Context Window Size". "Availability" indicates whether the model can be hosted on-premise (Local) or if an API call to an externally managed server is needed (API). For those that offer both, we highlight the method we used for the study in bold. "License Type" refers to whether the model weights are publicly available (Open) or not (Proprietary). 
\begin{table}[h!]
    \centering
    \resizebox{\columnwidth}{!}{
        \renewcommand{\arraystretch}{1.2} 
        \begin{tabular}{|l|c|c|c|c|}
            \hline
            \textbf{Model Name} & \textbf{Number of} &
            \textbf{Context Window} & 
            \textbf{Availability} & 
            \textbf{License Type} \\
            \textbf{} & \textbf{Parameters} & \textbf{Size (in tokens)} & \textbf{} & \textbf{} \\
            \hline
            Artigenz-Coder-DS-6.7B & 6.7B & 16,384 & Local/\textbf{API} & Open \\
            \hline
            CodeQwen1.5-7B-Chat & 7B & 64K & Local/\textbf{API} & Open \\
            \hline
            DeepSeek-V2.5 & 236B & 128K & Local/\textbf{API} & Open \\
            \hline
            GPT-3.5-Turbo & Unknown & 4,096 & API & Proprietary \\
            \hline
            GPT-4 & Unknown & 8,192 & API & Proprietary \\
            \hline
        \end{tabular}
    }
    \vspace{1mm}
    \caption{Selected Models for Code Generation}
    \label{tab:model-info}
    \vspace{-10pt}
\end{table}


\subsubsection{Repository Selection}
To select open-source repositories using the serverless architectural style, we utilize the Wonderless \cite{eskandani2021wonderless} and AWSomePy \cite{giuseppe_awsomepy} datasets. The Wonderless dataset is a collection of 1,877 serverless applications from GitHub in various languages. 72.2\% of the repositories included are in JavaScript or TypeScript, 19\% in Python, and 2.7\% in Java. The AWSomePy dataset consists of 145 AWS Lambda based serverless functions written in Python. We aim to choose at least one repository in JavaScript, TypeScript and Python each as they make up almost 70\% of runtimes used on AWS Lambda \footnote{\url{https://www.datadoghq.com/state-of-serverless/}}. In our selection, we also seek to ensure that the repository is not a demo or toy application, by filtering based on number of GitHub stars and manual inspection. Finally, we desire repositories that have tests defined with high test coverage to perform evaluations related to $RQ_{1}$, as described in \ref{study:eval}. 

First, we sort the datasets based on GitHub stars and forks in decreasing order, to prioritize popular real world repositories. We then look for repositories that contain tests. As a first step heuristic, for every repository, we look for files or folders with "test" as a substring of their name. Then, for repositories with $\geq 30$ stars on GitHub, we manually evaluate the quantity and quality of tests present. We found that some repositories had test files that were empty or too trivial to justify their selection. Additionally, several repositories contained serverless functions as only a small part of the entire codebase, usually for testing, and did not have tests for them. We also reject repositories that have been archived, since they may be deprecated or no longer relevant or maintained. For those repositories that had an acceptable number and quality of testcases, we calculated the test coverage and results. Based on these results, we selected four repositories.
We describe the selected repositories in Table \ref{tab:repo-info} below. All the repositories selected were from the Wonderless dataset.
\begin{table}[h!]
    \centering
    \begin{tabular}{|l|c|c|c|c|}
        \hline
        \textbf{Repository Name} & \textbf{Language} & \textbf{Stars} & \textbf{Forks} & \textbf{No. of Functions} \\
        \hline
        codebox-npm & Javascript & 352 & 27 & 10 \\
        \hline
        laconia & Javascript & 326 & 30 & 15 \\
        \hline
        TagBot & Python & 91 & 18 & 2 \\
        \hline
        StackJanitor & Typescript & 37 & 2 & 5 \\
        \hline
    \end{tabular}
    \vspace{1mm}
    \caption{Selected Repositories}
    \label{tab:repo-info}
    \vspace{-15pt}
\end{table}

\subsubsection{Function Selection}
For each repository, we selected up to 3 functions to mask so that they can be generated by the LLM, enabling us to assess the model's performance across varying levels of complexity, length, and structure. We desire functions that have tests associated with them to enable the evaluation of the generated code. To this end, for each selected repository, we calculated the test coverage using the testing framework used in the repository. We then selected the functions with the highest statement coverage, using source lines of code as a tiebreaker. We finally select 10 serverless functions for masking across the 4 repositories.

\subsubsection{Context Selection} \label{study:context}
In our study, we experimented with different levels of detail in the context provided to the LLM for code generation, focusing on two primary sources: the README file and a summary of the codebase which includes the architectural information.

For the README-based context, we used the repository’s README file, which typically offers a concise overview, including a brief description, functionalities, and usage instructions. 
However, since README files often lack implementation details, this approach evaluated the LLM’s ability to generate code with minimal context.

For the codebase summary-based context, we masked the target serverless function to generate a detailed codebase summary that excluded it. 
We utilized Gemini-1.5-Pro (mentioned in \ref{study:llm-selection}) to extract the architectural structure and functionality by identifying all other functions, their paths, and corresponding code (shown in \ref{box:codebase_summary_prompt}).
This structured summary, enhanced with keywords like "components," "connectors," and "relationships", provided context for evaluating the LLM’s ability to reconstruct the masked function accurately.

\begin{tcolorbox}[colback=orange!5!white, colframe=orange!95!white, colbacktitle=orange!95!white, title=Prompt for Codebase Summarization]
\small
    \textbf{Role Definition}\\
    \textbf{Task Definition}\\
    Here is the codebase, with the path name of the file followed by the contents of the file in triple backticks:

    \textit{\{FUNCTION PATH 1\}}\\
    \texttt{```}\\
    \textit{\{FUNCTION CODE 1\}}\\
    \texttt{```}
    \textit{\{FUNCTION PATH 2\}}\\
    \texttt{```}\\
    \textit{\{FUNCTION CODE 2\}}\\
    \texttt{```}
    
    \vdots
    \vspace{-5pt}
\end{tcolorbox}
\normalsize
\begin{minipage}{\linewidth}
\label{box:codebase_summary_prompt}
\end{minipage}
\vspace*{-10pt}

Additionally, masking a function involves generating its function description (shown in \ref{box:func_desc_prompt}), which is later incorporated in the prompt for the LLM to generate the function from its description. We used Gemini-1.5-Pro to produce this function description from the function code. The full prompts can be viewed in our replication package.

\begin{tcolorbox}[colback=orange!5!white, colframe=orange!95!white, colbacktitle=orange!95!white, title=Prompt for Function Description Generation]
\small
    \textbf{Role Definition}\\
    \textbf{Task Definition}\\
    The function path and the function code itself (enclosed in triple backticks) are provided below:

    \textit{\{MASKED FUNCTION PATH\}}\\
    \texttt{```}\\
    \textit{\{MASKED FUNCTION CODE\}}\\
    \texttt{```}
    \vspace{-7pt}
\normalsize
\end{tcolorbox}
\begin{minipage}{\linewidth}
\label{box:func_desc_prompt}
\end{minipage}
\vspace*{-10pt}

\subsubsection{Generation Methods} \label{study:generation}
We follow multiple generation methods through different prompts\\
\textbf{Prompt Types}:
From the two kinds of context described above, we create three prompts for function generation:
\begin{enumerate}
    \item \textit{Zero Shot with README} (Type 1 Prompt): As shown in \ref{box:type1}, it contains no examples of other serverless functions and their descriptions and only provides the README file of the repository as context. The description of the masked function is provided and the model is tasked with generating the code for the serverless function.
    \begin{tcolorbox}[colback=orange!5!white, colframe=orange!95!white, colbacktitle=orange!95!white, title=Zero Shot with README, width=\linewidth]
\small
    \textbf{Role Definition}: You are a ...\\
    You are working with a FaaS codebase whose README is as follows:\\
    \textit{\{README\}}

    \tcblower 
\small
    \textbf{Task Definition}\\
    The function should have the following functionality:\\
    \textit{\{FUNCTION DESCRIPTION\}}
    
    \textbf{Detailed Instructions Including Formatting}
  \normalsize
\end{tcolorbox}
\begin{minipage}{\linewidth}
\label{box:type1}
\end{minipage}
    \vspace{-5pt}
    \item \textit{Zero Shot with Codebase Summarization} (Type 2 Prompt): As shown in \ref{box:type2}, it also contains no examples of other functions and descriptions, but includes the architectural information through the summary of the codebase, along with the description of the masked serverless function.
    \begin{tcolorbox}[colback=orange!5!white, colframe=orange!95!white, colbacktitle=orange!95!white, title=Zero Shot with Codebase Summarization]
\small
    \textbf{Role Definition}: You are a ...\\
    You are working with a FaaS codebase...:\\
    \textit{\{CODEBASE SUMMARY\}}

    \tcblower
\small
    \textbf{Task Definition}\\
    The function should have the following functionality:
    
    \textit{\{FUNCTION DESCRIPTION\}}
    
    \textbf{Detailed Instructions Including Formatting}
\normalsize
\end{tcolorbox}
\begin{minipage}{\linewidth}
\label{box:type2}
\end{minipage}

    \item \textit{Few Shot with Codebase Summarization} (Type 3 Prompt): Along with the architectural information through the summary of the codebase, this prompt (shown in \ref{box:type3}) also contains descriptions and function code of other serverless functions in the repository. These serve as guides for the model to generate the code for the masked function from the given description. This few-shot prompt allows the LLM to learn in-context without modifying model parameters, as described in Section \ref{backg:llm_prompting}.
    \begin{tcolorbox}[colback=orange!5!white, colframe=orange!95!white, colbacktitle=orange!95!white, title=Few Shot with Codebase Summarization]
\small
    \textbf{Role Definition}: You are a ...\\
    You are working with a FaaS codebase...:\\
    \textit{\{CODEBASE SUMMARY\}}\\

    \textbf{Task Definition}\\
    Here are some examples. \\
    \textit{\{EXAMPLE FUNCTION DESCRIPTION 1\}}\\
    \texttt{```}\\
    \textit{\{EXAMPLE FUNCTION CODE 1\}}\\
    \texttt{```}\\
    \textit{\{EXAMPLE FUNCTION DESCRIPTION 2\}}\\
    \texttt{```}\\
    \textit{\{EXAMPLE FUNCTION CODE 2\}}\\
    \texttt{```}

    \tcblower
\small
    The function you generate should have the following functionality:
    
    \textit{\{FUNCTION DESCRIPTION\}}\\
    \textbf{Detailed Instructions Including Formatting}
\normalsize
\end{tcolorbox}
\begin{minipage}{\linewidth}
\label{box:type3}
\end{minipage}
\end{enumerate}
The full prompts can be found in our \href{https://doi.org/10.5281/zenodo.14539782}{replication package}. When making the prompts, we made an effort to be as detailed and specific as possible.

\begin{figure}[t]
    \centering
    \includegraphics[width=0.9\linewidth]{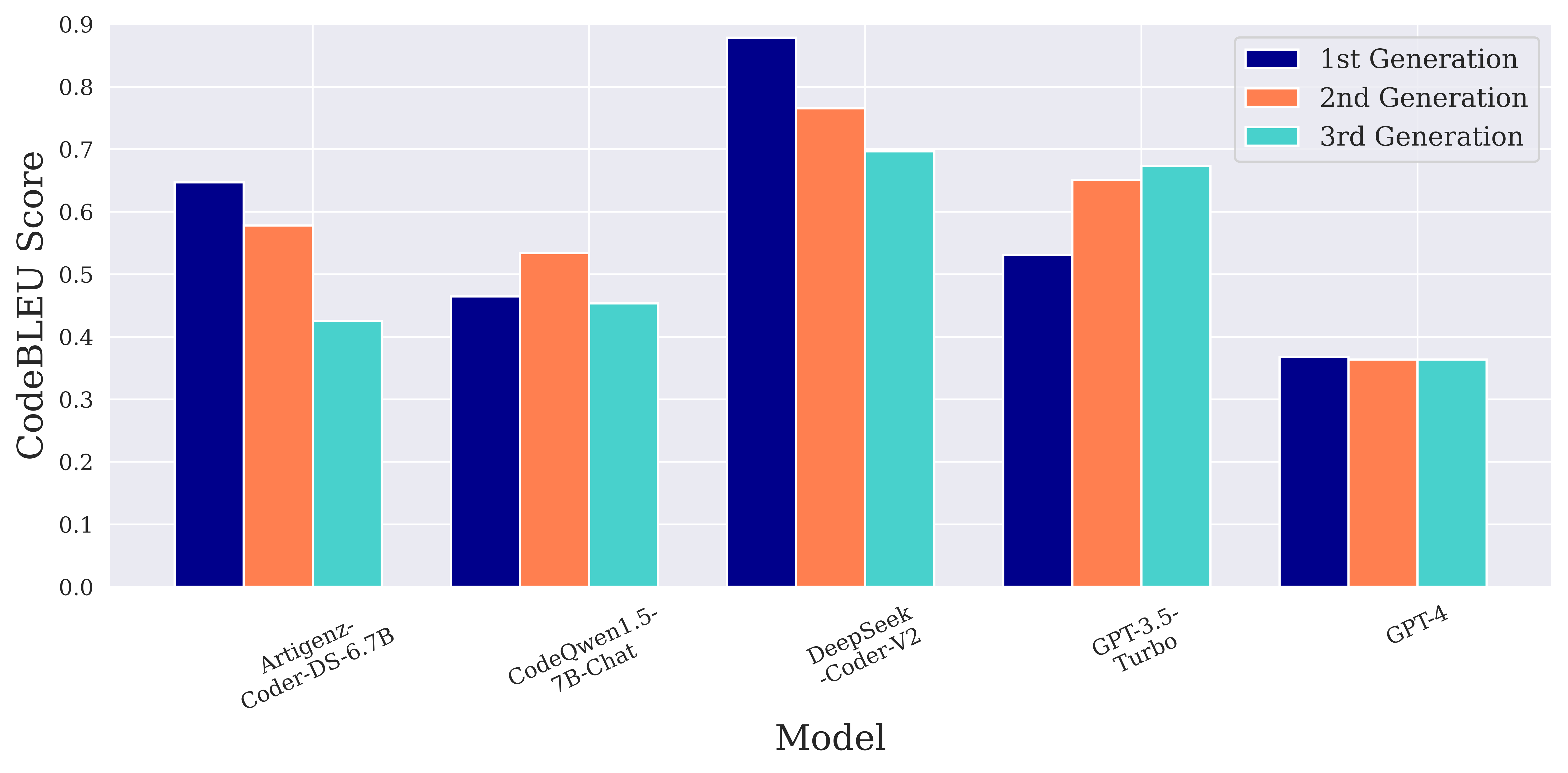}
    \vspace{-9pt}
    \caption{Avg. Pairwise CodeBLEU Scores for Generated Functions per Model}
    \label{fig:codebleu-consistency}
    \vspace{-15pt}
\end{figure}

\noindent
\textbf{Consistency Check}: LLMs are probabilistic, and can produce different outputs even with the same prompt. In the context of our work, this can result in the generated code passing tests and being of good quality in one run, but not in the other. To alleviate this issue, we verify if multiple generations using the same context generate similar code. We compare code similarity using CodeBLEU \cite{ren2020codebleu}, which is a weighted combination of n-gram matches\footnote{\scriptsize an n-gram match is $n$ tokens matching in order between the input and the output}, syntactic matches in the Abstract Syntax Trees (ASTs) and semantic data-flow matches. The results 
are shown in Figure \ref{fig:codebleu-consistency}. We see that generated functions are quite similar to each other for a model, and conclusions drawn by evaluating one function generated by a model will hold even when the model is given multiple tries.
We finally generate 145 serverless functions for evaluation.



\subsubsection{Evaluation Metrics} \label{study:eval}
We perform three kinds of evaluations on the LLM generated serverless functions:\\
\textit{Functional Correctness Through Testing}: To address $RQ_{1}$, we evaluated both the original and generated code using the existing tests in each repository. Specifically, we recorded the number of passing and failing tests for the entire codebase as well as for the individual functions selected for generation. After generating each function, we re-ran the tests and recorded the updated counts of passing and failing cases.

To address $RQ_{1.1}$, we initially conducted this evaluation without any code modifications. Here, we used the generated response of the LLM, cleaned to include only the code. Although this process was done manually, it can be automated by extracting only the text within triple backticks (\texttt{\textasciigrave \textasciigrave \textasciigrave}), as the model was instructed to provide code in this format. However, manual cleaning was necessary for cases where the model did not follow the formatting instructions.

To address $RQ_{1.2}$, we identified and fixed code generation errors made by LLMs. Song et al.\cite{song2023empirical} study the errors in LLM-generated code and create a taxonomy of observed errors. We filter this to only include those errors that were frequently observed and solvable with minimal human intervention, shown in Table \ref{table:llm_error_table}.
On average, we spent 15 minutes per function fixing these errors. Following these corrections, we recorded the updated counts of passing and failing tests, both for the overall codebase and individual functions.

\begin{table}[h!]
    \centering
    \scriptsize
    \resizebox{\columnwidth}{!}{
        \begin{tabular}{|p{2cm}|p{2.5cm}|p{3cm}|}
            \hline
            \textbf{Error Categories} & \multicolumn{2}{c|}{\textbf{Error Types}} \\
            \hline
            \multirow{6}{*}{Semantic Errors} 
            & Condition Error
            & Missing condition \\
            & & Incorrect condition \\
            \cline{2-3}
            & Constant Value Error
            & Constant value error \\
            \cline{2-3}
            & Reference Error
            & Wrong method/variable \\
            &  &Undefined name \\
            \cline{2-3}
            & Calculation Error
            & Incorrect arithmetic/ comparison operation \\
            \cline{2-3}
            & Incomplete Code
            & Missing one statement \\
            \cline{2-3}
            & Memory Error
            & Infinite loop \\
            & & Integer overflow \\
            \hline
            \multirow{7}{*}{Syntactic Errors} 
            & Conditional Error
            & If error \\
            \cline{2-3}
            & Loop Error
            & For/ While error \\
            \cline{2-3}
            & Return Error
            & Incorrect return value \\
            \cline{2-3}
            & Method Call Error
            & Incorrect function name/ arguments \\
            & & Incorrect method call target \\
            \cline{2-3}
            & Assignment Error
            & Incorrect arithmetic \\
            & & Incorrect constant \\
            & & Incorrect variable name \\
            & & Incorrect comparison \\
            \cline{2-3}
            & Import Error
            & Import error \\
            \hline
        \end{tabular}
        }
    \vspace{1mm}
    \caption{Errors Identified and Fixed in Generated Function}
    \label{table:llm_error_table}
    \vspace{-10pt}
\end{table}

\textit{Code Quality through Code Metrics}: Though the component generated may be functional, in the sense that it passes tests, we also desire code that is of good quality, to ensure quality attributes such as maintainability, readability and extensibility. In addressing $RQ_{2}$, we quantify code quality using code level metrics.
Our choice of metrics encompasses those that measure size (through Source Lines of Code (SLOC) and Halstead's Measure), testing difficulty (McCabe's Complexity) and reading difficulty (cognitive complexity). 
Our choice for metrics is guided by previous work by Jin et al. \cite{jin_metrics}, who analyze Java, JavaScript, and Python repositories. The popular Chidamber \& Kemerer object-oriented metrics \cite{chidamber1994metrics} are not applicable, since serverless functions written in the languages studied in this work need not be object-oriented. This is further supported by Jin et al. \cite{jin_metrics}, who also avoid using OO metrics in similar contexts. We calculate the following metrics for the original and generated serverless functions: 
\begin{enumerate}
    \item \textit{Source Lines of Code (SLOC)} quantifies the size of a program by counting the number of lines which contain source code.
    \item \textit{Halstead's Software Science Metric (Volume)} \cite{halstead_measure} is another measure to quantify the size of a program which considers programs as a collection of tokens (operators and operands).
    \item \textit{McCabe's Cyclomatic Complexity (CC)} \cite{mccabe1976complexity} is used to measure the complexity of a program's control flow. It provides insight into the testability of the program, since one with a high cyclomatic complexity has more paths that need to be tested.
    \item \textit{Cognitive Complexity (CogC)} \cite{campbell2018cognitivecomplexity} is a metric designed to measure the understandability of code.
\end{enumerate}

\textit{Code Similarity using CodeBLEU}: In answering $RQ_{2}$, we also measure how syntactically similar LLM generated serverless functions are to human written ones through the CodeBLEU \cite{ren2020codebleu} metric, which is described in \ref{study:generation}. A high CodeBLEU score indicates that the human-written and generated code are very similar, and the models may have appropriately created the summary and generated function. Though CodeBLEU measures semantic data-flow similarity, we utilize it to compare the syntactic similarity of serverless functions and use testing to evaluate semantic correctness. This is because the generated function may not pass tests despite a high CodeBLEU score with the original function due to subtle differences in the code that affects functionality or creates errors but does not significantly reduce CodeBLEU, such as incorrect imports or exchanged variables. \\

\vspace{-6pt}
\section{Results} \label{results}
\begin{table*}[htb!]
    \centering
    \scriptsize
    \renewcommand{\arraystretch}{1.2} 
    \setlength{\tabcolsep}{3pt} 
    \begin{tabular}{|l|l|c|c|c|c|c|c|c|c|c|c|c|c|c|c|c|}
        \hline
        \multirow{5}{*}{\textbf{Model Name}} & \multirow{5}{*}{\textbf{Repository Name}} & \multicolumn{14}{c|}{\textbf{Percentage of Test Cases Passed (\%)}} \\
        \cline{3-16}
        & & \multicolumn{2}{c|}{\textbf{Original}} & \multicolumn{12}{c|}{\textbf{Generated}} \\
        \cline{3-16}
        & & & & \multicolumn{4}{c|}{\textbf{Type 1 Prompt}} & \multicolumn{4}{c|}{\textbf{Type 2 Prompt}} & \multicolumn{4}{c|}{\textbf{Type 3 Prompt}} \\
        \cline{5-16}
        & \textbf{(Language)} & \textbf{\faGithub} & \textbf{\textcolor{orange}{$\lambda$}} & \multicolumn{2}{c|}{\textbf{\faGithub}} & \multicolumn{2}{c|}{\textbf{\textcolor{orange}{$\lambda$}}} & \multicolumn{2}{c|}{\textbf{\faGithub}} & \multicolumn{2}{c|}{\textbf{\textcolor{orange}{$\lambda$}}} & \multicolumn{2}{c|}{\textbf{\faGithub}} & \multicolumn{2}{c|}{\textbf{\textcolor{orange}{$\lambda$}}} \\
        \cline{3-16}
        & & \textbf{\faUser\textcolor{red}{\faTimes}} & \textbf{\faUser\textcolor{green}{\faCheck}} & \textbf{\faUser\textcolor{red}{\faTimes}} & \textbf{\faUser\textcolor{green}{\faCheck}} & \textbf{\faUser\textcolor{red}{\faTimes}} & \textbf{\faUser\textcolor{green}{\faCheck}} & \textbf{\faUser\textcolor{red}{\faTimes}} & \textbf{\faUser\textcolor{green}{\faCheck}} & \textbf{\faUser\textcolor{red}{\faTimes}} & \textbf{\faUser\textcolor{green}{\faCheck}} & \textbf{\faUser\textcolor{red}{\faTimes}} & \textbf{\faUser\textcolor{green}{\faCheck}} & \textbf{\faUser\textcolor{red}{\faTimes}} & \textbf{\faUser\textcolor{green}{\faCheck}} \\
        \hline
        \multirow{5}{*}{Artigenz-Coder-DS-6.7B} & laconia (JS) & 100 & 100 & 74.36 & 74.36 & 0 & 0 & 74.36 & 74.36 & 0 & 0 & 74.36 & 74.36 & 0 & 0 \\
        \cline{2-16}
        & codebox-npm (JS) & 100 & 100 & 28.94 & 60.07 & 0 & 0 & 28.94 & 91.94 & 0 & 0 & 60.07 & 93.41 & 0 & 0 \\
        \cline{2-16}
        & StackJanitor (TS) & 100 & 100 & 67.57 & 67.57 & 0 & 0 & 67.57 & 67.57 & 0 & 0 & 67.57 & 67.57 & 0 & 0 \\
        \cline{2-16}
        & Tagbot (PY) & 93.33 & 100 & 0 & 0 & 0 & 0 & 82.67 & 82.67 & 0 & 0 & - & - & - & -\\
        \cline{2-16} 
        & \textbf{Avg. Model Performance} & \textbf{98}	& \textbf{100} & \textbf{43} & \textbf{51} & \textbf{0} & \textbf{0} & \textbf{63} & \textbf{58} & \textbf{0} & \textbf{0} & \textit{\textbf{67}} & \textit{\textbf{74}} & \textbf{0} & \textbf{0}\\
        \hline
        \multirow{5}{*}{CodeQwen1.5-7B-Chat} & laconia (JS) & 100 & 100 & 74.36 & 74.36 & 0 & 0 & 74.36 & 74.36 & 0 & 0 & 74.36 & 74.36 & 0 & 0 \\
        \cline{2-16}
        & codebox-npm (JS) & 100 & 100 & 28.94 & 60.07 & 0 & 0 & 28.94 & 93.04 & 0 & 16.67 & 64.1 & 95.6 & 22.22 & 66.67 \\
        \cline{2-16}
        & StackJanitor (TS) & 100 & 100 & 68.47 & 68.47 & 11.11 & 11.11 & 67.57 & 70.27 & 0 & 33.33 & 67.57 & 70.27 & 0 & 33.33 \\
        \cline{2-16}
        & Tagbot (PY) & 93.33 & 100 & 0 & 0 & 0 & 0 & 82.67 & 86.67 & 0 & 37.5 & - & - & - & -\\
        \cline{2-16} 
        & \textbf{Avg. Model Performance} & \textbf{98} & \textbf{100} & \textbf{43} & \textbf{51} & \textbf{3} & \textbf{3} & \textbf{63} & \textbf{81} & \textbf{0} & \textbf{22} & \textit{\textbf{69}} & \textit{\textbf{80}} & \textbf{7} & \textbf{33}\\
        \hline
        \multirow{5}{*}{DeepSeek-Coder-V2} & laconia (JS) & 100 & 100 & 74.36 & 74.36 & 0 & 0 & 74.36 & 74.36 & 0 & 0 & 74.36 & 79.49 & 0 & 66.67 \\
        \cline{2-16}
        & codebox-npm (JS) & 100 & 100 & 31.87 & 91.94 & 0 & 0 & 91.94 & 94.51 & 0 & 41.67 & 94.14 & 95.6 & 38.89 & 66.67 \\
        \cline{2-16}
        & StackJanitor (TS) & 100 & 100 & 67.57 & 81.08 & 0 & 33.33 & 67.57 & 90.09 & 0 & 62.5 & 75.68 & 95.5 & 0 & 79.17 \\
        \cline{2-16}
        & Tagbot (PY) & 93.33 & 100 & 82.67 & 86.67 & 0 & 37.5 & 82.67 & 90.67 & 0 & 50 & - & - & - & -\\
        \cline{2-16} 
        & \textbf{Avg. Model Performance}  & \textbf{98} & \textbf{100} & \textbf{64} & \textbf{84} & \textbf{0} & \textbf{18} & \textbf{79} & \textbf{87} & \textbf{0} & \textbf{39} & \textit{\textbf{81}} & \textit{\textbf{90}} & \textit{\textbf{13}} & \textit{\textbf{71}}\\
        \hline
        \multirow{5}{*}{GPT-3.5-Turbo} & laconia (JS) & 100 & 100 & 74.36 & 74.36 & 0 & 0 & 74.36 & 76.92 & 0 & 33.33 & 74.36 & 94.87 & 0 & 62.5 \\
        \cline{2-16}
        & codebox-npm (JS) & 100 & 100 & 60.81 & 91.94 & 0 & 0 & 60.81 & 94.87 & 0 & 55.56 & 60.81 & 95.6 & 11.11 & 66.67 \\
        \cline{2-16}
        & StackJanitor (TS) & 100 & 100 & 67.57 & 78.38 & 0 & 0 & 67.57 & 76.58 & 0 & 62.5 & 67.57 & 76.58 & 0 & 62.5 \\
        \cline{2-16}
        & Tagbot (PY) & 93.33 & 100 & 0 & 84 & 0 & 12.5 & 82.67 & 90.67 & 0 & 50 & - & - & - & -\\
        \cline{2-16}
        & \textbf{Avg. Model Performance} & \textbf{98} & \textbf{100} & \textbf{51} & \textbf{82} & \textbf{0} & \textbf{3} & \textbf{71} & \textbf{85} & \textbf{0} & \textbf{50} & \textit{\textbf{68}} & \textit{\textbf{89}} & \textit{\textbf{4}} & \textbf{64}\\
        \hline
        \multirow{4}{*}{GPT-4} & laconia (JS) & 100 & 100 & 74.36 & 76.92 & 0 & 33.33 & 74.36 & 100 & 0 & 100 & 74.36 & 87.18 & 0 & 20.83 \\
        \cline{2-16}
        & codebox-npm (JS) & 100 & 100 & 31.87 & 91.94 & 0 & 0 & 60.81 & 62.27 & 0 & 33.33 & 93.77 & 95.6 & 30.56 & 66.67 \\
        \cline{2-16}
        & StackJanitor (TS) & 100 & 100 & 67.57 & 75.68 & 0 & 0 & 67.57 & 90.09 & 0 & 62.5 & 67.57 & 86.49 & 0 & 62.5 \\
        \cline{2-16}
        & Tagbot (PY) & 93.33 & 100 & 82.67 & 86.67 & 0 & 37.5 & 82.67 & 90.67 & 0 & 50 & - & - & - & -\\
        \cline{2-16}
        & \textbf{Avg. Model Performance} & \textbf{98} & \textbf{100} & \textbf{43} & \textbf{83} & \textbf{0} & \textbf{24} & \textbf{51} & \textbf{86} & \textbf{0} & \textbf{61} & \textit{\textbf{79}} & \textit{\textbf{90}} & \textit{\textbf{10}} & \textit{\textbf{50}}\\
        \hline
        \multicolumn{2}{|c|}{\textbf{Average}} & \textbf{98} & \textbf{100} & \textit{\textbf{49}} & \textit{\textbf{70}} & \textbf{1} & \textit{\textbf{10}} & \textit{\textbf{66}} & \textit{\textbf{79}} & \textbf{0} & \textit{\textbf{34}} & \textit{\textbf{73}} & \textit{\textbf{85}} & \textit{\textbf{7}} & \textit{\textbf{44}}\\
        \hline
    \end{tabular}
    \vspace{2mm}
    \caption{Test Case Pass Rates: \faIcon{github} for Codebase Tests, \textbf{\textcolor{orange}{$\lambda$}} for Function Tests; \faUser\textcolor{red}{\faTimes}(No Human Intervention) vs. \faUser\textcolor{green}{\faCheck}(With Human Intervention)}
    \label{table:RQ1-table}
    \vspace{-18pt}
\end{table*}

\begin{figure}
    \centering
    \includegraphics[width=0.9\linewidth]{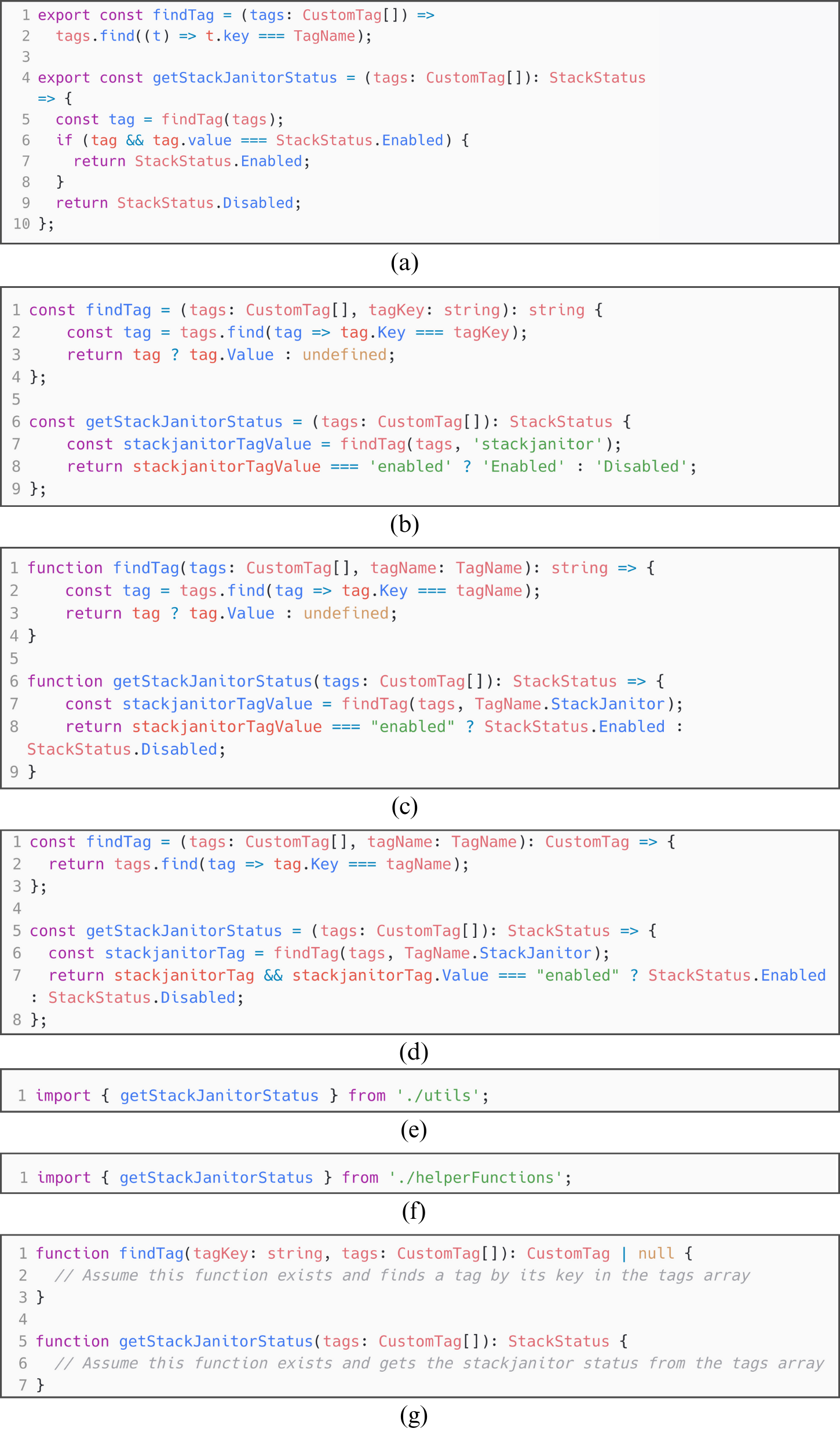}
    \caption{Snippets from Functions: (a) Original, (b-d) DeepSeek-Coder-V2 with Type 1-3 prompts, (e-g) Artigenz-Coder-DS-6.7B with Type 1-3 prompts.}
    \label{fig:examples}
    \vspace{-15pt}
\end{figure}

\label{sec:rq1-1}
\textbf{RQ}$_{1.1}$: Can LLMs generate functional serverless functions without human intervention?

Table \ref{table:RQ1-table} summarizes the percentage of test cases passed for the entire codebase and individual serverless functions, comparing original and LLM-generated functions across all models and prompt types. Type 3 prompts achieved the highest average test passing rates,  with 73\% of tests passed for the entire codebase. Comparing Type 1 prompt and Type 2 prompt for serverless function generation, the latter resulted in a higher test passing rates. 

Figures \ref{fig:examples}(b-d) and {}(e-f) provide examples illustrating how the prompt types impact the generated function's structure and functionality. In Figure \ref{fig:examples}(b), generated by DeepSeek-Coder-V2 with the type 1 prompt, \texttt{findTag()} and \texttt{getStackJanitorStatus()} rely on hard-coded strings (e.g., \texttt{"stackjanitor"} and \texttt{"enabled"}). This approach can reduce flexibility and misalign the function with actual codebase expectations. In contrast, Figure \ref{fig:examples}(c), generated by the same model with the type 2 prompt, avoids hard-coded strings by referencing a variable, showing an improvement. However, this version still lacks complete logic. Figure \ref{fig:examples}(d) shows a further improvement from the type 3 prompt, where \texttt{findTag()} is defined to return the tag object itself, which aligns closer to the original serverless function (in Figure \ref{fig:examples}(a)) and demonstrates the benefit of more detailed context.

Table \ref{table:RQ1-table} reveals that larger models namely DeepSeek-Coder-V2, GPT-3.5-Turbo, and GPT-4 significantly outperformed smaller models (CodeQwen1.5-7B-Chat and Artigenz-Coder-DS-6.7B), regardless of prompt type. Notably, DeepSeek-Coder-V2 achieved the highest performance without human intervention, passing over 81\% of codebase tests and 13\% of individual function tests with type 3 prompt. GPT-4 followed with 79   \% and 10\% test passing rates, respectively. In contrast, smaller models such as Artigenz-Coder-DS-6.7B and CodeQwen1.5-7B-Chat showed significantly lower passing rates. Figure \ref{fig:examples}(d) generated by DeepSeek-Coder-V2, and Figure \ref{fig:examples}(g) generated by Artigenz-Coder-DS-6.7B with the type 3 prompt, illustrate this difference - the former provides an accurate response in line with the original function, while the latter produces only a stub.  

The choice of language also influenced the functionality, with JavaScript-based repositories achieved higher pass rates in comparison to the other languages.  

\hfill\\
\textbf{RQ}$_{1.2}$: Can LLMs generate functional serverless functions with minimal human intervention?

The results after applying minimal human intervention to the generated functions, as shown in Table \ref{table:RQ1-table}, demonstrate a significant increase in test passing rates across all models and prompt types. DeepSeek-Coder-V2, which performed well with respect to RQ1.1, achieved over 90\% success on codebase tests and 71\% on individual function tests when using the Type 3 prompt. GPT-4 and GPT-3.5-Turbo also showed substantial performance improvements, with pass rates close to DeepSeek-Coder-V2. Even the lower-performing models, Artigenz-Coder-DS-6.7B and CodeQwen1.5-7B-Chat, saw marked  improvements: their codebase test pass rates increased from 67\% to 74\% and 69\% to 80\%, respectively. However, their results still fell short of the larger models. \\
For instance, as illustrated in Figure \ref{fig:examples}(d) — an output generated by DeepSeek-Coder-V2 — correcting a return error based on the guidance in Table \ref{table:llm_error_table}, specifically by adjusting \texttt{findTag()} to return the entire tag object instead of just the tag value, improved the functionality of the generated code. Conversely, Figure \ref{fig:examples}(g) shows an output generated by Artigenz-Coder-DS-6.7B with the Type 3 prompt, which is a stub to benefit from minor corrections, highlighting the limitations of smaller models in producing functional code even with minimal human intervention.

\begin{figure*}[h]
    \centering
    \includegraphics[width=0.95\textwidth]{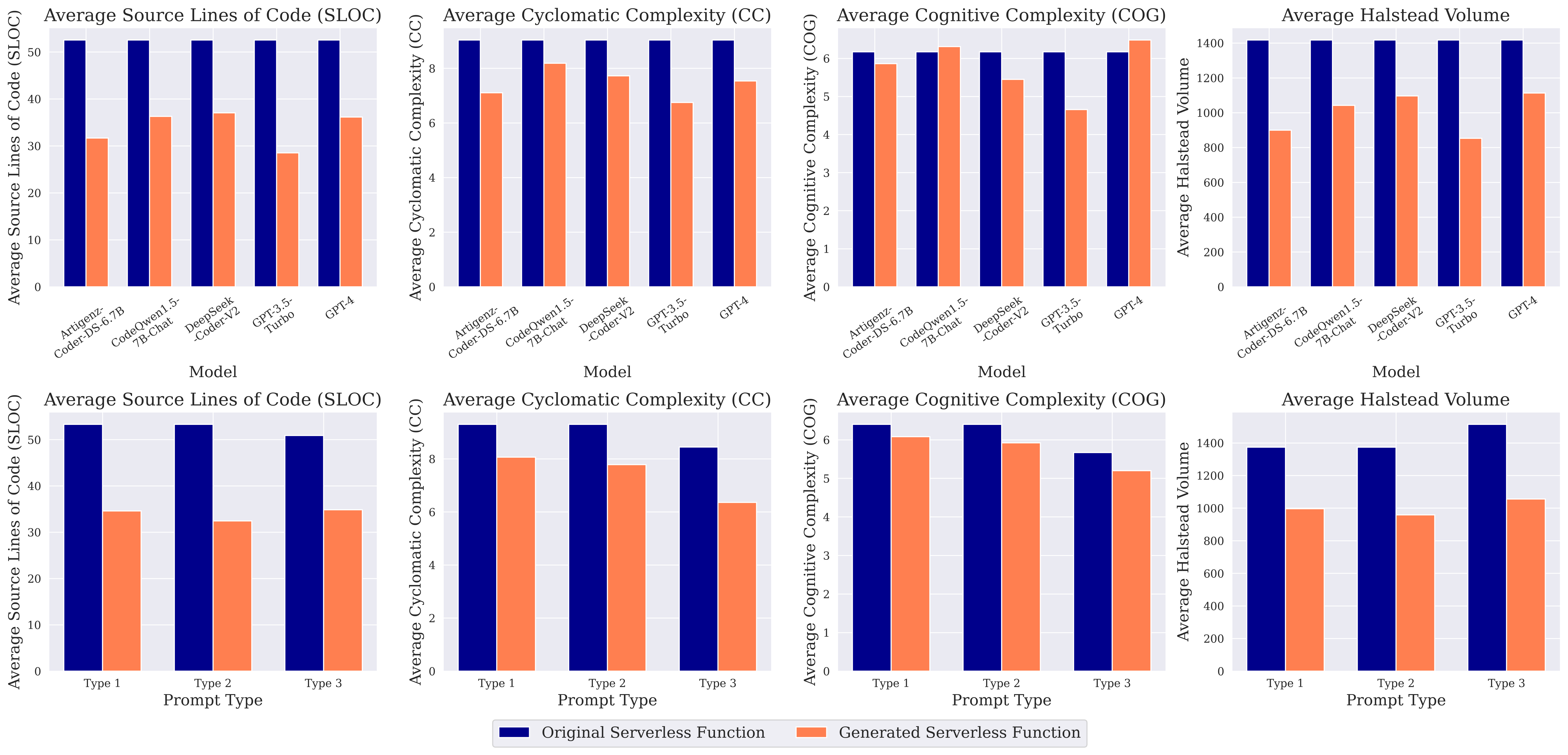}
    \caption{Code Quality Metrics of Original and Generated Functions per Model (first row), per Prompt Type (second row)}
    \label{fig:metrics-graphs}
    \vspace{-10pt}
\end{figure*}

\textbf{RQ}$_2$: How does the code quality of LLM-written serverless functions compare to human-written code?
From Figure \ref{fig:metrics-graphs}, we see an overall reduction in the values of all code metrics except for some outliers in CogC for LLM generated serverless functions. For average SLOC, we see a reduction for all models and prompt types. Interestingly, we see that coding specific models generate more SLOC than the general purpose models, despite Artigenz-Coder-DS-6.7B and CodeQwen1.5-7B-Chat being much smaller than GPT-3.5-Turbo and GPT-4. Average CC of the functions generated by the models is also lower than the original, with coding specific models again demonstrating higher values despite smaller size. We observe that CodeQwen-1.5-7B-Chat and GPT-4 generated functions with higher CogC than the original functions, while Artigenz-Coder-DS-6.7B, DeepSeekCoder-V2 and GPT-3.5-Turbo have lower values than the original functions, with GPT-3.5-Turbo generating the most understandable code. Halstead volume of generated functions were much lower than the original functions across all models.  

Regarding the difference in code quality that the different amount of context provided in the prompt to the models make, we find no appreciable difference in average SLOC and average Halstead Volume. However, we see a marked reduction in average CC and average CogC when more context is provided to the model, with the lowest values demonstrated by the Few Shot Prompt with Codebase Summarization. 

\begin{figure}[h]
    \centering
    \includegraphics[width=0.85\linewidth]{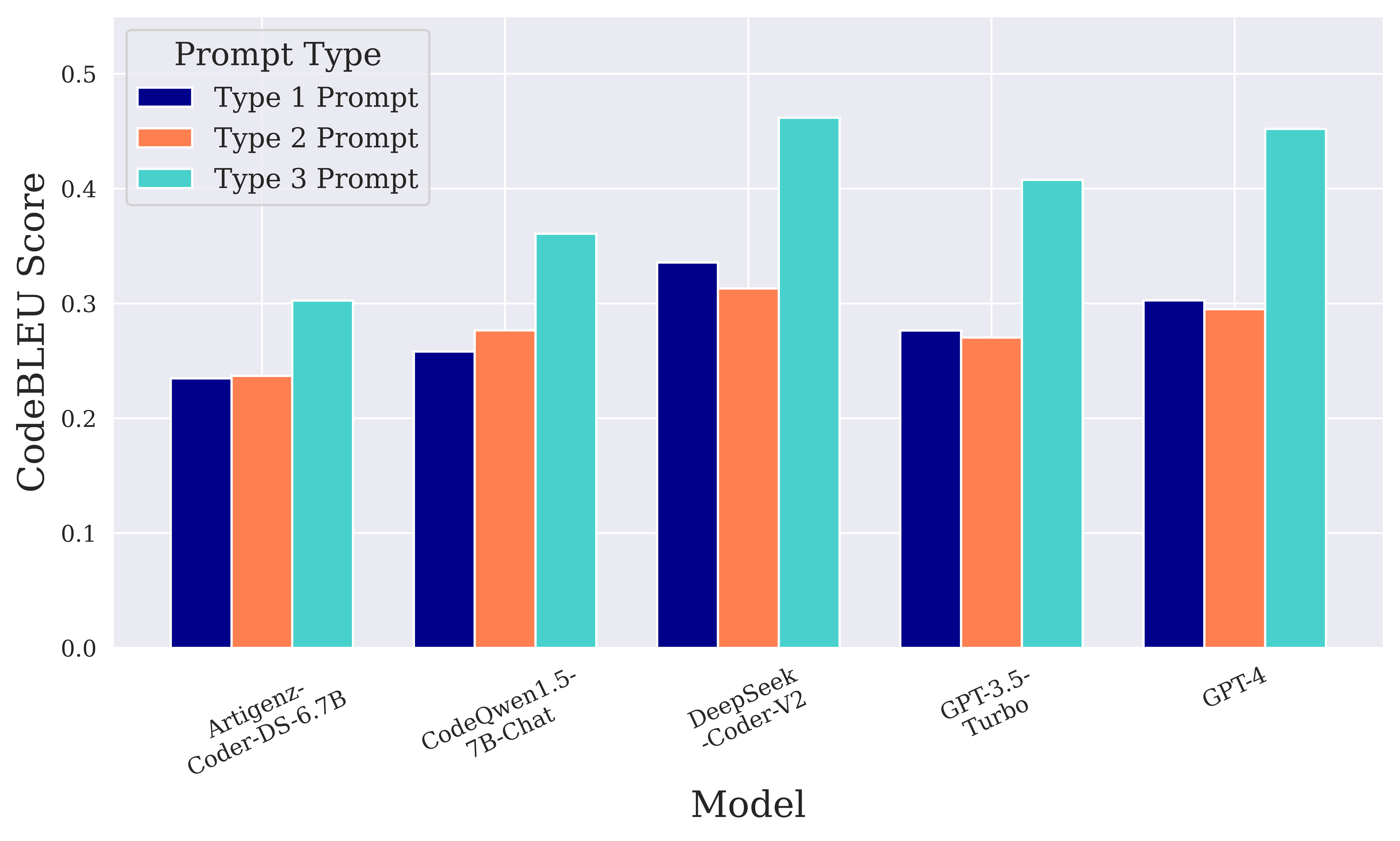}
    \vspace{-5pt}
    \caption{Avg. CodeBLEU Scores for Generated Functions per Model and Prompt Type}
    \label{fig:codebleu-models}
    \vspace{-20pt}
\end{figure}

On comparing the similarity of human-written and LLM-generated functions using CodeBLEU score, we see from Figure \ref{fig:codebleu-models} that Type 3 prompts make all models generate functions that are much more similar to the human written functions. We also see a correlation where larger models generate code more similar to the originals, with DeepSeek-Coder-V2 having the highest similarity score. Interestingly, in many cases, providing the codebase summarization in Type 2 prompt reduced CodeBLEU score compared to using the Type 1 prompt, though the former produced functions that passed more tests.

\newtcolorbox{mybox}{colback=orange!5!white,colframe=orange!75!black}
\newtcolorbox{myboxtitle}[1]{colback=orange!5!white,colframe=orange!75!black,fonttitle=\bfseries,title=#1}

\section{Discussion} \label{discussion}
In this section, we discuss lessons learnt from our study and threats to validity.

\subsection{Can LLMs generate functional serverless functions? ($RQ_1$)}
Yes, LLMs can be used to generate functional serverless functions with varying performance based on model and prompt type, but not entirely autonomously.

We observed that different levels of detail in the context provided to the LLM for serverless function generation significantly impact its performance. The Type 3 prompt produces the most functional serverless functions. This prompt type
containing comprehensive context along with description and function code of other serverless functions in the repository, gave the models a better understanding of the codebase structure and function interdependencies
enabled better understanding of codebase structure and interdependencies, which was particularly beneficial to the larger models - DeepSeek-Coder-V2, GPT-4, and GPT-3.5-Turbo. This underlines the importance of both model scale and context in generating functional serverless functions. Smaller models - Artigenz-Coder-DS-6.7B and CodeQwen1.5-7B-Chat, constrained by their size, tend to generalize more, reducing the precision needed for function generation.

DeepSeek-Coder-V2 demonstrated the best overall results, across multiple functions and prompt types without human intervention. GPT-4 and GPT-3.5-Turbo also showed strong performance, though GPT-4 surpassed GPT-3.5-Turbo in adhering to the function requirements. 

Interestingly, with respect to RQ1.2, involving minimal human intervention, the smaller models required substantial corrections, while the larger models needed relatively minor adjustments. However, in certain cases, the improvement in the functionality of the generated code was more pronounced for GPT-3.5-Turbo than GPT-4 which led GPT-3.5-Turbo to outperform GPT-4. Upon manual inspection of the generated functions, we found that GPT-3.5-Turbo tends to produce simpler initial outputs which, although often lacking full functionality, are generally easier to refine than GPT-4's more complex responses. GPT-4, while thorough in adhering closely to function requirements, occasionally produces code that is more challenging to adjust with minor changes.

As observed in \ref{sec:rq1-1}, JS-based repositories achieve higher pass rates, likely due to its prevalence in serverless functions. This insight can be used when selecting LLMs for architectural component generation.

Overall, while LLMs, especially the larger models, showcase the potential to generate functional serverless functions with minimal human intervention, there remains a gap between generated and fully functional human-written architectural components. This necessitates further research on leveraging LLMs to reach human-level proficiency in generating serverless functions autonomously, with an emphasis on maintaining a human in the loop to refine functionality, handle complex logic, and ensure practical applicability.

\begin{tcolorbox}[colback=orange!5!white, colframe=orange!95!white, colbacktitle=orange!95!white]
\small
    \textbf{Main Findings for RQ1}: LLMs can generate functional serverless functions with varying performance based on model and prompt type, but not to the extent to which humans can. Nonetheless, this method can be employed to aid architectural component generation.
\end{tcolorbox}

\subsection{How does the code quality of LLM-written serverless functions compare to human-written code? ($RQ_{2}$)}

LLM-generated serverless functions show a reduction in 
Source Lines of Code, Cyclomatic Complexity, and Halstead Volume 
compared to human-written code. This suggests that LLMs produce concise and less complex code that is easier to understand and maintain. However, this simplicity can sometimes come at the expense of functionality, as demonstrated in Figure \ref{fig:examples}(g), where essential logic was omitted, impacting functional accuracy. The code generation models Artigenz-Coder-DS-6.7B and CodeQwen1.5-7B-Chat produced more SLOC and higher CC than general-purpose models like GPT-4 and GPT-3.5-Turbo, despite their smaller size. Among the models, larger ones, particularly DeepSeekCoder-V2, achieved the highest CodeBLEU scores, highlighting the importance of model size and prompt detail.

When comparing prompt types, Type 3 prompt, which provides detailed architectural and functional context, significantly reduced CC and CogC, leading to simpler, and more understandable code. It also produced high similarity between LLM-generated and human-written functions, as illustrated in Figures \ref{fig:examples}(a) and (d). Despite the improvements in complexity metrics, Type 3 prompts had minimal impact on SLOC and Halstead Volume, indicating that while context simplifies logic, it does not necessarily shorten code length. This observation highlights the nuanced influence of context in shaping code quality.

An interesting anomaly was observed with Type 2 prompts, which occasionally reduced CodeBLEU scores despite generating functions that passed more tests compared to Type 1. This suggests that functional accuracy doesn’t always align with human-code similarity. 

Overall, while LLMs generate simpler and more understandable code, their limited ability to handle complex logic autonomously points to a trade-off between code quality and functionality, necessitating further refinement to achieve the desired functionality.

\begin{tcolorbox}[colback=orange!5!white, colframe=orange!95!white, colbacktitle=orange!95!white]
\small
    \textbf{Main Findings for RQ2}: LLMs can generate good quality serverless functions compared to human-written code. Larger models, combined with prompts that capture architectural and functional context, produce simpler, more understandable, and functionally accurate code, albeit with some trade-offs in complexity and functionality.
\end{tcolorbox}

\subsection{Defining Software Architecture in the Age of GenAI}
Over the years, numerous definitions of software architecture (SA) have been proposed over the years, including by Perry and Wolf \cite{perrywolf_architecture}, Garlan and Shaw \cite{garlan1993_component_connector} and Jansen and Bosch \cite{bosch_designdecisions}.
Refer \cite{What_is_your_definition_of_software_architecture_2010} for more definitions.
When one considers component generation as a black-box, Jansen and Bosch's definition of SA as a set of design decisions \cite{bosch_designdecisions} is more pertinent, since the architect only makes design decisions that are developed automatically.
However, when moving deeper, during generation, it is necessary to consider the \textit{components} that already exist in the system, how they are \textit{connected} and what \textit{constraints} exist, which is Garlan and Shaw's definition\cite{garlan1993_component_connector}. As we move into an age where GenAI is increasingly used in SA, it is imperative to be able to move between these definitions depending on the level of abstraction used.
The need of the hour is accurate and detailed documentation of these design decisions to enable LLMs to generate the components and connectors subject to the constraints defined in the design decisions. These design decisions can be complemented by organizational context, and all of this can in turn benefit from the usage of GenAI, such as for documenting Architectural Decision Records (ADRs) as proposed by Dhar et al.\cite{dhar_adr}.

\subsection{The Path Forward for GenAI in SA} \label{path-forward}
Perhaps the biggest challenge we faced in conducting this study was finding high quality data. Despite using published datasets, we found the quality of tests and components to be mostly unsatisfactory. Machine learning models need voluminous high quality data \cite{jain_data_quality_ml}, and while this data exists to an extent for language modeling and code snippet generation, it is much more scarce when considering architectural component generation. There currently do not exist architecture specific tasks and benchmarks to evaluate LLMs, probably due to the aforementioned data scarcity. \\
The next issue that needs to be solved is the specific approach to using LLMs for generating architectural components. Various methods like Chain-of-thought prompting \cite{wei2022chainofthought}, retrieval-augmented-generation \cite{lewis2020rag}, knowledge graphs \cite{pan2024llmkg} \cite{abdelaziz2021codekg}, and agentic frameworks \cite{huang2023agentcoder, zhuge2024agentasajudge} exist, but their effectiveness for software architecture specific tasks have not been explored. LLMs could also be seen as a solution to address the learning curve of ADLs/DSLs where natural language could be converted to an architectural model and further code could be generated using model transformations. However, addressing these problems requires deeper collaboration between the SA and NLP communities. 

\subsection{Threats to Validity}
We follow the categorization provided by Wohlin et al. \cite{wohlin2012experimentation} and also provide brief explanation about efforts taken to mitigate the identified threats or why it is not possible, as suggested by Verdecchia et al. \cite{verdecchia_threats}. \\
    \textbf{External Validity}: Selection of LLMs used and their generation parameters (such as temperature) for creating the context in the form of the codebase summary and for generating the serverless function poses a threat to external validity. However, we systematically select diverse LLMs using peer-reviewed published leaderboards, namely ChatBot-Arena \cite{chiang2024chatbotarenaopenplatform} and EvalPlus \cite{liu2024code_evalplus}, with EvalPlus specifically evaluating the coding abilities of LLMs. 
    An interesting threat to external validity can be contamination of LLMs \cite{dong2024contamination}, which is when LLMs are trained or fine-tuned on data that is later used to evaluate their performance, leading to unrealistically high scores. Since most LLMs are not open about the data they are trained on, we have no way to mitigate this issue, especially since the repositories listed in Wonderless and AWSomePy were released in the public domain on which LLMs could have been trained. Despite this possibility, as shown in the results in Section \ref{results}, LLMs were not able to generate completely functional serverless functions. \\
    \textbf{Internal Validity}: 
    The effectiveness of the tests in the repositories selected can be a threat to internal validity. However, we manually selected repositories which had high test coverage to mitigate this. 
    Selection of metrics presents another threat, since both code quality and similarity are quite abstract concepts. To address this, we used metrics commonly used in the SE and NLP communities for measuring the same. \\
    \textbf{Construct Validity}: 
    A threat to construct validity stems from the selection of repositories and serverless functions in them. However, we utilize published datasets - Wonderless \cite{eskandani2021wonderless} and AWSomePy \cite{giuseppe_awsomepy}. We filter them on the number of stars, quantity and quality of tests and test coverage to avoid picking demo projects and retain repositories and serverless functions that have real-world relevance. 

\section{Conclusion and Future Work} \label{conclusion}
This study, whose code and data we make publicly available, \footnote{Code and data available at: \url{https://doi.org/10.5281/zenodo.14539782}} seeks to empirically explore the capabilities of Large Language Models to automatically generate software architectural components. We do this in context of serverless functions which is an event-driven architectural stlye, due to the small size of their basic architectural unit. Using a range of models, including general-purpose models like GPT-4 and code generation models like DeepSeekCoder-V2, and diverse prompt types, we evaluate the generated architectural components for both functional correctness and code quality. We find that while LLMs often fail to generate fully functional serverless functions autonomously, they can serve as a valuable starting point, as minimal human intervention significantly improves their functionality, enabling them to pass more tests. We also observe that LLM  generated serverless functions display better metrics related to code quality, such as cyclomatic complexity, and cognitive complexity. Furthermore, in its current state, one cannot, and probably should not completely remove the human from the component generation process, suggesting a human-centered GenAI approach, as put forth by the Copenhagen Manifesto \cite{copenhagen_manifesto}.\\
Future work involves exploring other techniques for generation used at the code level, such as those mentioned in \ref{path-forward}, including exploring Retrieval Augmented Generation, multi-agent frameworks, and other architectural styles such as microservices and monoliths with larger components.


\bibliographystyle{ieeetr}
\bibliography{main}

\end{document}